\documentclass[aps,prl,superscriptaddress]{revtex4}

\newcommand{\m}{\mathbf}

\newcommand{\pdsmall}[2]{{\partial #1}/{\partial #2}}

\usepackage{graphicx}
\usepackage{amsmath}
\usepackage{verbatim}
\usepackage{color}

\begin{document}

%\preprint{}

\title{Optomechanics in an ultrahigh-$Q$ slotted 2D photonic crystal cavity}


\author{Amir H. Safavi-Naeini}
\thanks{These authors contributed equally to this work.}
\author{Thiago P. Mayer Alegre}
\thanks{These authors contributed equally to this work.}
\author{Martin Winger}
\author{Oskar Painter}
\email{opainter@caltech.edu}
\affiliation{Thomas J. Watson, Sr., Laboratory of Applied Physics, California Institute of Technology, Pasadena, CA 91125}
\date{\today}

\begin{abstract}
Optical guided wave dielectric structures with nanoscale slots are known to provide strong electric field enhancements in the slotted region, enabling strong light-matter interactions.  Here we demonstrate an ultrahigh-$Q$ slotted 2D photonic crystal cavity for the purpose of obtaining strong interaction between the internal light field and the mechanical motion of the slotted structure.  The experimentally measured optical quality factor is $Q=1.2\times 10^6$ for a cavity with a simulated effective optical modal volume of $V_{\text{eff}}=0.04$~$(\lambda)^3$.  Optical transduction of the thermal Brownian motion of the fundamental in-plane mechanical resonance of the slotted structure ($\nu_{m}=151$~MHz) is performed, from which an optomechanical coupling of $g^*_{\text{OM}}/2\pi=140$~GHz/nm is inferred for an effective motional mass of $m_{\text{eff}}=20$~pg.  Dynamical back-action of the optical field on the mechanical motion, resulting in both cooling and amplication of the mechanical motion, is also demonstrated.   
\end{abstract}

\pacs{}

\maketitle

%\section{Introduction}

The strength of the interaction between light and matter, which is fundamental to many applications in nonlinear and quantum optics, depends on the ability to create a large optical energy density, either through increased photon number or photon localization.  This may be achieved by creating optical cavities with large quality factors $Q$ and simultaneously small modal volumes $V_\text{eff}$. The mode volume $V_\text{eff}$ in particular can be decreased through the introduction of slots, increasing the electric field intensity in low-index regions of the device. As such, slotted photonic crystal cavities~\cite{Robinson2005} and waveguides~\cite{Riboli2007} have been previously proposed and applied to create highly sensitive detectors of motion~\cite{Eichenfield2009a,Lin2009} and molecules~\cite{DiFalco2009}. They have also more recently been studied in the context of Purcell enhancement of spontaneous emission from embedded quantum dots~\cite{Gao2010}.

In the canonical optomechanical system, consisting of a Fabry-Perot resonator with an oscillating end-mirror~\cite{Kippenberg2008}, the radiation pressure force per cavity photon is given by $\hbar g_\text{OM}\equiv \hbar\pdsmall{\omega_o}{x}=\hbar \omega_{o}/L_{\text{OM}}$, where $\omega_{o}$ is the cavity resonance frequency, $x$ is the position of the end mirror, and $L_\text{OM}$ is approximately equal to the physical length of the cavity.  The optomechanical coupling length in such ``diffraction-limited'' cavities is limited approximately by the wavelength of light, $L_{\text{OM}}\gtrsim \lambda$.  Large optomechanical coupling, approaching $g_{\text{OM}} = \omega_{o}/\lambda$, has recently been realized in several different guided wave optical cavity geometries utilizing nanoscale slots~\cite{Eichenfield2009a,Lin2009,Roh2010}. In this work we design, fabricate, and measure the optomechanical properties of a slotted two-dimensional (2D) photonic crystal cavity formed in a Silicon membrane. Due to the strong optical confinement provided by a sub-$100$~nm slot and a two-dimensional photonic bandgap, this cavity structure is demonstrated to have an optical quality factor $Q > 10^6$, a sub-wavelength $L_\text{OM}$, and a deep sub-cubic-wavelength optical mode volume.

%\section{Design}

\begin{figure}[btp]
\includegraphics[width=0.85\columnwidth]{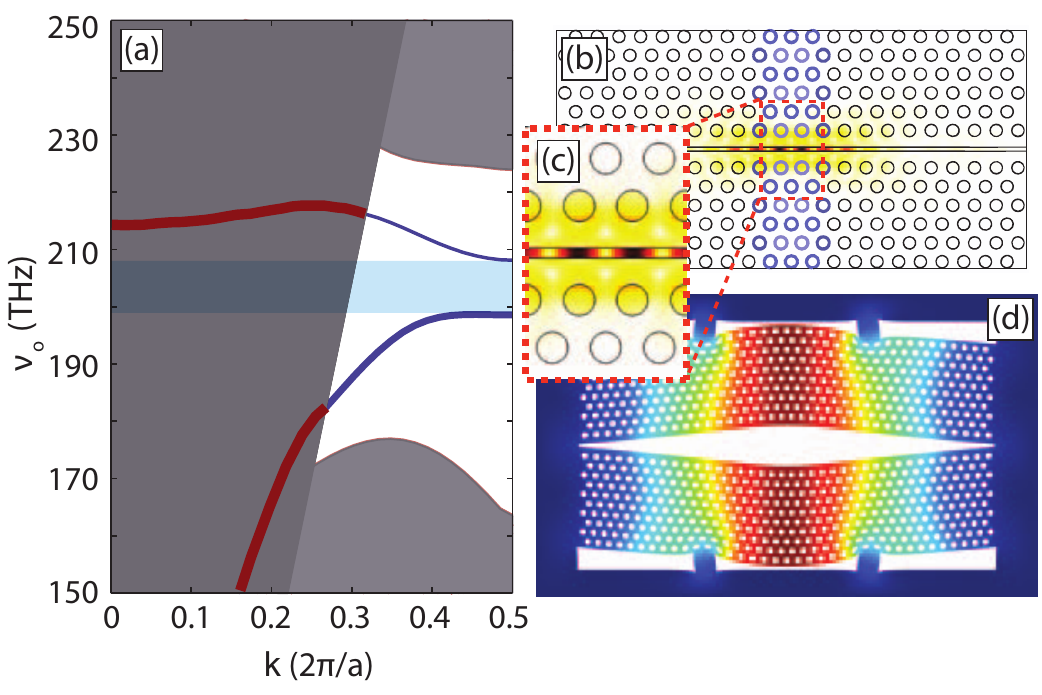}
\caption{(a) Band diagram for a slotted W1 waveguide formed in a thin ($t=220$~nm) Silicon layer. The waveguide slot size is $s = 0.2 a$, with lattice hole radius $r = 0.285 a$ for a nominal lattice constant of $a = 470~\text{nm}$. The light gray shade indicates the guided mode continua, while the dark gray represents the unguided continua of radiation modes. The blue curve is the waveguide band used to create the heterostructure cavity. Leaky resonant modes are shown by the thick red line. The photonic quasi-bandgap (for TE-like modes of even vector parity) is highlighted in blue. (b) Electric field intensity, $|\m E(\m r)|^2$ of the optical mode. (c) Zoom-in of the slotted region showing strong optical field confinement. (d) Total displacement field $|\m Q(\m r)|$ of the simulated fundamental mechanical mode at $146.1~\text{MHz}$.}\label{fig:simulated}
\end{figure}

A common approach to forming photonic crystal optical circuits is to etch a pattern of holes into a thin dielectric film such as the top Silicon device layer in a Silicon-On-Insulator (SOI) microchip.  An effective means of forming resonant cavities in such quasi-2D slab photonic crystal structures is to weakly modulate the properties of a line-defect waveguide~\cite{ref:Johnson4,ref:Chutinan2000,Song2005}. Applying this same design principle to slotted photonic crystal waveguides~\cite{DiFalco2008}, optical cavities with $Q\leq 5\times 10^4$ have been experimentally demonstrated~\cite{Gao2010,DiFalco2009}. A major source of optical loss in real fabricated structures is light scattering out of the plane of the slab.  One class of optical states which play an important role in determining scattering loss are the resonant leaky modes of the slab. These optical resonances are localized to the slab and yet have wavevector components which radiate energy into the surrounding cladding.  To reduce the effects of these modes it is preferable to engineer a structure where the photonic crystal waveguide has no leaky mode bands crossing the localized cavity mode frequency.  For the popular W1 waveguide~\cite{ref:Johnson4,ref:Chutinan2000} with a slot added in the waveguide center, we have found that the choice of the slot width is crucial to avoiding coupling to leaky resonances.  Figure~\ref{fig:simulated}(a) shows the bandstructure of a slotted W1 waveguide with a hole radius $r = 0.285a = 134$~nm, slot size $s = 0.2a = 94$~nm, thickness $t=220$~nm, and nominal lattice constant of $a=470$~nm.  A large bandgap for both guided and leaky modes (a ``quasi-bandgap'') is clearly present in this structure for the TE-like (even vector parity) modes of the waveguide.  On the otherhand, for slot widths $s > 0.25 a$ the quasi-bandgap of the waveguide closes due to the presence of leaky resonant bands. 

% Although these resonances lie within the ``light cone'' of the cladding surrounding the slab, and thus are typically ignored when considering the photonic bandgap properties of a quasi-2D crystal, there presence in close proximity to the frequency of a localized cavity mode can lead to substantial radiation leakage from defects in a fabricated structure. 

In order to form a localized cavity resonance, we begin with the slotted cavity waveguide structure of Figure~\ref{fig:simulated}(a).  A localized resonance is created from the lower frequency waveguide band by reducing smoothly the local lattice constant from a nominal value of $a=470~\text{nm}$ to a value of $a=450~\text{nm}$ in the center of the cavity.  Three-dimensional finite-element-method (FEM) simulations of the optical and mechanical properties of the resulting cavity structure were performed.  The simulated electric field intensity of the fundamental confined optical mode is shown in Fig.~\ref{fig:simulated}(b).  This mode has a resonance wavelength of $\lambda_{o}\approx1550$~nm, a theoretical radiation-limited $Q > 10^6$ and an effective optical mode volume of $V_\text{eff} = 0.04$~$(\lambda_{o})^3$. 

To allow for mechanical motion of the structure, three rectangular holes of dimensions $4.9~\mu \text{m} \times 1.0~\mu\text{m}$ are cut on each side of cavity device as shown in Fig.~\ref{fig:simulated}(d). FEM simulations show that this allows for a fundamental in-plane mechanical mode of motion with frequency $\omega_{m}/2\pi=146.1~\text{MHz}$ and an effective motional mass of $m_{\text{eff}}=20~\text{pg}$. The optomechanical coupling between the localized optical and mechanical modes is computed using a variation~\cite{Johnson2002} of the Hellman-Feynman perturbation theory adopted for optomechanical systems~\cite{Eichenfield2009d}, yielding a sub-wavelength optomechanical coupling length of $L_\text{OM} = 400~\text{nm}$.  This optomechanical coupling length corresponds to an optical frequency shift per nanometer of mechanical displacement of $g_\text{OM} = 2\pi\times480~\text{GHz/nm}$, or in the quantum realm a phonon-photon coupling rate of $g = g_\text{OM} \sqrt{\hbar/2 m_\text{eff} \omega_m} = 2\pi \times 800~\text{kHz}$.

\begin{figure}[btp]
\includegraphics[width=0.85\columnwidth]{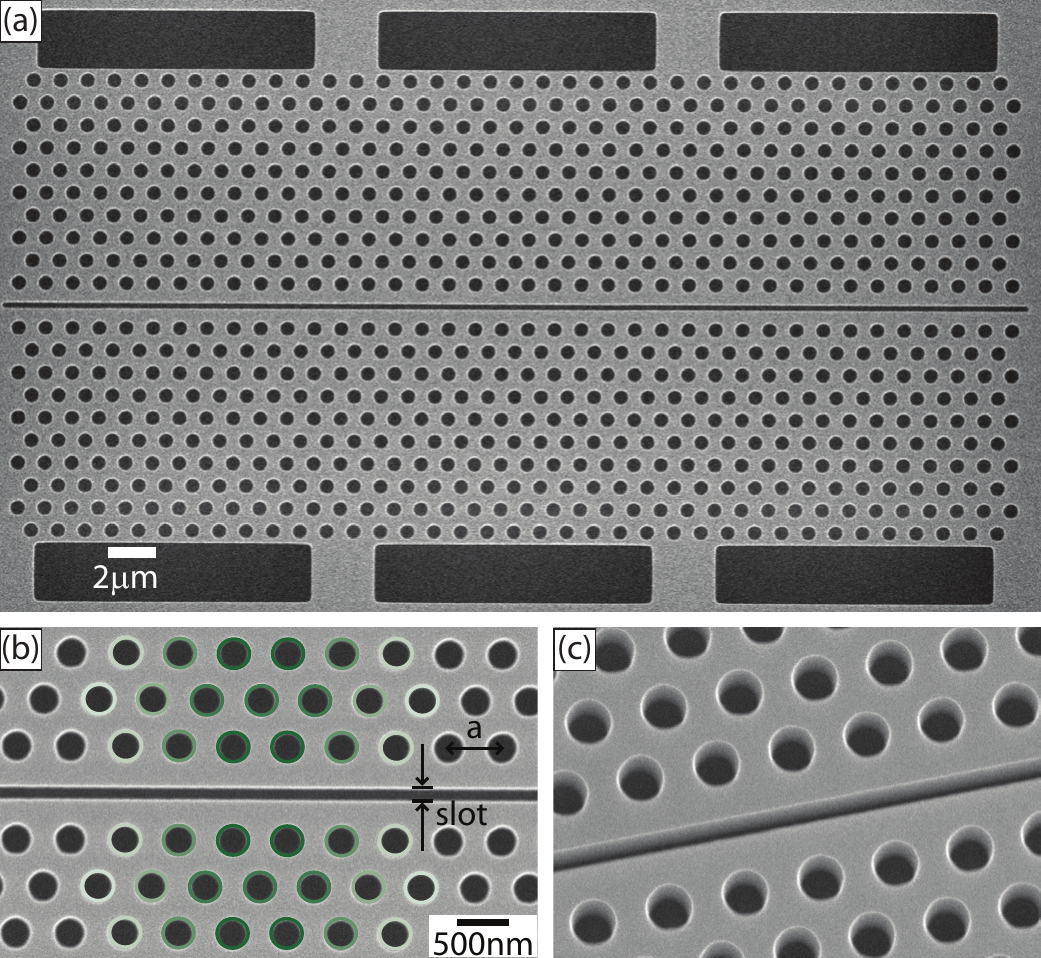}
\caption{(a) SEM image of the fabricated sample. (b) Zoom-in SEM image of the cavity region, with the heterostructure defect cavity region highlighted in false color, and (c) SEM image showing the etched sidewalls of the slot and holes. \label{fig:sem}}
\end{figure}

Slotted cavities with the dimensions stated above are fabricated using a Silicon-On-Insulator wafer from SOITEC ($\rho=4$-$20$~$\Omega\cdot$cm, device layer thickness $t=220$~nm, buried-oxide layer thickness $2$~$\mu$m). The cavity geometry is defined by electron beam lithography followed by reactive-ion etching to transfer the pattern through the $220~\text{nm}$ silicon device layer. The cavities are undercut using $\text{HF:H}_2\text{O}$ solution to remove the buried oxide layer, and cleaned using a piranha/HF cycle~\cite{Borselli2006}. A scanning electron microscope (SEM) micrograph of a final device is shown in Fig.\ref{fig:sem}(a). Fig.~\ref{fig:sem}(b) and Fig.~\ref{fig:sem}(c) show the local waveguide defect and slotted region of the cavity, respectively.

\begin{figure}[btp]
\includegraphics[width=0.85\columnwidth]{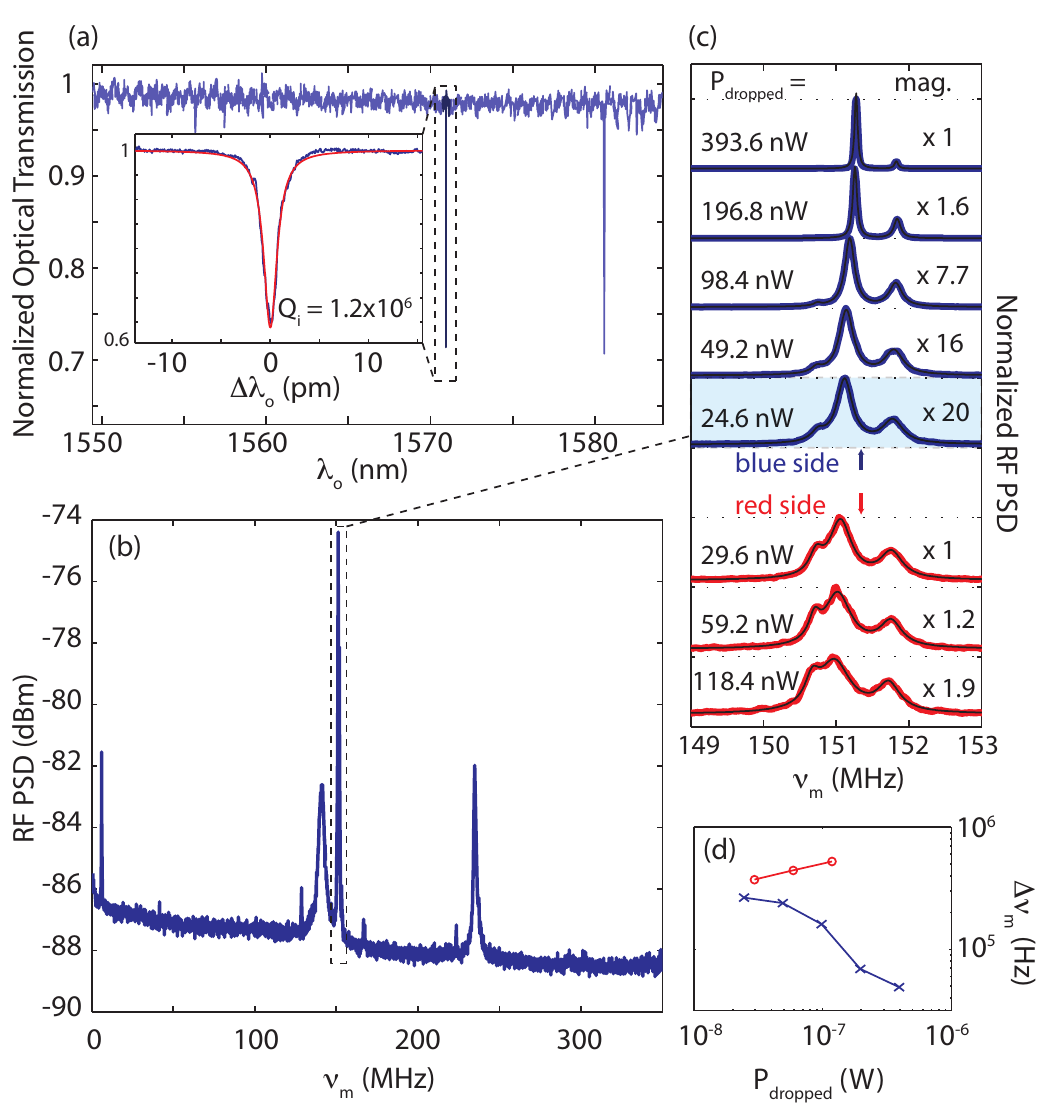}
\caption{(a) Normalized optical transmission spectrum showing the first and second order optical cavity modes. (inset) Transmission spectrum for the first order mode showing an intrinsic quality factor of $Q_i=1.2\times10^6$. (b) RF-PSD of the photodetected signal, indicating a series of resonance peaks corresponding to mechanical motion of the patterned slab. (c) RF-PSD around the frequency of the fundamental in-plane mechanical resonance for various optical powers dropped into the cavity.  A pair of optical attenuators are used, one at the input and one at the output of the cavity, so that the optical power reaching the photodetector is constant and the measured RF-PSD is only dependent upon the amplitude of mechanical motion.  The red (bottom three) and blue (top five) spectra represent spectra taken with red and blue input laser detuning from the cavity resonance, respectively.  Denoted for each spectrum is the optical power dropped into the cavity and a scale factor used to normalize the peak height in each spectrum to a common value. (d) Linewidth of the fundamental in-plane mechanical mode extracted from the spectra in (c).  $\times = $ blue detuning, $\circ = $ red detuning.} \label{fig:measured} 
\end{figure}

The resulting devices are characterized optically using a swept-wavelength external-cavity laser ($\lambda = 1510-1590~\text{nm}$, $\Delta\lambda < 300~\text{kHz}$), connected to a dimpled fiber-taper probe~\cite{Michael2007}. A broadband cavity transmission spectrum is shown in Fig.~\ref{fig:measured}(a), with the first- and second-order optical cavity modes separated by roughly by $10~\text{nm}$, in agreement with simulations.  For the first-order mode, optical $Q$ on the order of $10^6$ is measured consistently in these devices.  A narrowband optical transmission spectrum (calibrated using a fiber Mach-Zender interferometer) for one such device is shown in the inset of Fig.~\ref{fig:measured}(a), with a measured instrinsic optical $Q_i = 1.2\times10^6$. 

The mechanical properties of the slotted photonic crystal cavity are measured by driving the system with the laser frequency locked to a detuning of a half-linewidth (blue or red) from the cavity resonance.  The transmitted cavity laser light is sent through an erbium doped fiber amplifier and then onto a high-speed photodetector.  The photodetected signal is sent to an oscilloscope ($2$~GHz bandwidth) where the electronic power spectral density (PSD) is computed. An example of the measured RF-spectrum from a typical slotted cavity device is shown in Fig.~\ref{fig:measured}(b). The fundamental in-plane mode, corresponding to the largest peak in the spectrum, is found to occur at a frequency of $\omega_{m}/2\pi=151~\text{MHz}$, very close to the simulated value of $146$~MHz. RF spectra for various dropped optical powers into the cavity are shown in Fig.~\ref{fig:measured}(c).  The corresponding mechanical linewidth is plotted in Fig.~\ref{fig:measured}(d).  The effects of the retarded component of the dynamical back-action~\cite{Kippenberg2008} of the light field on the mechanical resonance are clear in both plots, with red (blue) detuning resulting in a reduction (amplification) in the mechanical resonance peak height and a broadening (narrowing) of the mechanical linewidth.  One curious aspect of the measured mechanical spectra, however, are the two smaller side peaks on either side of the main resonance line.  These side peaks are thought to arise due to a mixing of the in-plane mode with nearby flexural (out-of-plane) resonances of the slab.  FEM simulations indicate the presence of several flexural modes located within a few MHz of the fundamental in-plane mode, which can mix with in-plane motion through a breaking of the vertical symmetry of the slab. As an example, SEM images show that the membranes are subject to weak stress-induced bowing, which can lead to this effect.  

The optomechanical coupling of the fundamental in-plane mechanical resonance can be estimated using two different methods.  The first method involves calibration of the optical powers and electronic detection system, and uses the fact that the transduced thermal Brownian motion of the mechanical resonator is proportional to $g^2_\text{OM}/m_{\text{eff}}$~\cite{Eichenfield2009}.  The second method compares the ratio of the RF power in the first and second harmonic of the mechanical frequency.  This method is independent of the absolute optical power and detection efficiency, and relies only on accurate knowledge of the optical linewidth.  Both of these methods were found to yield an experimental optomechanical coupling of $g^\ast_\text{OM} = 2\pi \times140~\text{GHz/nm}$~($L_{\text{OM}}=1.4$~$\mu$m) for the fundamental in-plane mechanical resonance~(assuming $m_{\text{eff}}=20$~pg), roughly a factor of $3.4$ times smaller than the FEM-estimated value.  As alluded to above, this discrepancy is thought to result from mixing of in-plane motion with optomechanically ``dark'' (i.e., no coupling to first-order in motional amplitude) flexural modes of the patterned slab.  

In summary, the slotted photonic crystal cavity described here reduces optical scattering loss through the avoidance of resonant leaky modes of the structure while simultaneously allowing for large electric field enhancement in the cavity slot region.  The demonstrated optical loss rate of the cavity is $\kappa/2\pi\approx 160$~MHz, which in conjunction with the high mechanical frequency ($\omega_{m}/2\pi=151$~MHz) of the fundamental in-plane mechanical resonance, puts this system in the resolved sideband limit of cavity optomechanics ($\kappa/2\omega_{m} < 1$).  The resolved sideband limit is important for a variety of applications, including optical cooling of the mechanical motion to the quantum mechanical ground-state~\cite{Marquardt2007,Wilson-Rae2007}.  The optomechanical coupling is estimated to be $g^\ast_\text{OM}/2\pi=140$~GHz/nm for the slotted cavity, due largely to the electric field enhancement in the slot and corresponding to one of the largest values measured to date~\cite{VanThourhout2010}.  The estimated $Q/V_\text{eff}$ ratio for the measured devices is $3\times10^7 (\lambda)^{-3}$, indicating that these slotted cavities may also find use in other applications such as Silicon-based cavity-QED~\cite{Gao2010} and sensing~\cite{DiFalco2009}. 

This work was supported by the DARPA/MTO ORCHID program through a grant from AFOSR, and the Kavli Nanoscience Institute at Caltech. ASN gratefully acknowledges support from NSERC.

%\bibliography{../slotted}

\end{document}